# Revisiting the Frictional Control of the Antarctic Circumpolar Current From the Energy Diagram

Takuro Matsuta[a], Yuki Tanaka[b] and Atsushi Kubokawa[a]

[a] Faculty of Environmental Earth Science, Hokkaido University*, Sapporo, Japan*

[b] Department of Ocean Sciences, Tokyo University of Marine Science and Technology, *Minato-ku, Japan*

*Corresponding author*: Takuro Matsuta, matsuta@ees.hokudai.ac.jp

ABSTRACT

The transport of the Antarctic Circumpolar Current (ACC) has been shown to increase with friction. Previous studies explained this counter-intuitive relationship called *frictional control* based on the eddy geometric parametrizations. They focused on the eddy momentum transfer and eddy energetics. To maintain the balance between wind stress and eddy interfacial form stress, eddy energy must remain unchanged as friction increases; this requires enhanced baroclinicity to compensate for stronger eddy energy dissipation. However, the independence of eddy energy has not been fully verified, and this interpretation assumes negligible barotropic energy conversion. To address this gap, we conduct sensitivity experiments in an idealized stratified reentrant channel with varying linear bottom drag. Numerical simulations show that eddy energy changes substantially with friction. Furthermore, in the high-drag regime, baroclinic energy conversion dominates eddy energy generation, whereas in the low-drag regime barotropic energy conversion contributes substantially. Despite these differences, baroclinicity increases with eddy energy dissipation across all regimes, although the relationship is somewhat weak in the low-drag regime owing to barotropic energy conversion. To explain this phenomenon, we extend the frictional control framework based on the Lorenz energy cycle. A simple scaling argument leads to a generalized frictional control, $s \sim D(E)/\tau_w$, where $s$ is baroclinicity, $D(E)$ is eddy energy dissipation, and $\tau_w$ is wind stress. This framework provides a natural extension of the existing framework and successfully explains the numerical results. These results indicate that eddy dissipation controls the baroclinicity; therefore, properly parameterizing the eddy dissipation rate is essential for representing ACC dynamics in ocean models.

## 1. Introduction

Ocean circulation is driven by external forcing, including wind stress, heat flux, and tidal mixing. Energy budget studies suggest that the adiabatic component of ocean circulation is sustained by wind stress (von Storch et al. 2012; Zemskova et al. 2015). Wind stress drives upwelling and enhances the isopycnal slope, which tends to increase available potential energy (APE) (Roquet et al. 2011; Aiki et al. 2011). The APE gained is released by baroclinic energy conversion. The Antarctic Circumpolar Current (ACC) is one of the hotspots of baroclinic energy conversion since there exist strong westerlies in the Southern Ocean. Because of this importance, the ACC energy cycle has been the subject of extensive research.

Eddy energy in the ACC has been discussed in the framework of the Lorenz energy cycle (LEC) (Chen et al. 2014; Youngs et al. 2017; Wu et al. 2017; Matsuta and Masumoto 2023; Matsuta et al. 2024). The LEC estimates energy conversions among mean kinetic energy (MKE), mean available potential energy (MAPE), eddy available potential energy (EAPE), and eddy kinetic energy (EKE) to quantify eddy-mean flow interactions. Baroclinic energy conversion links the MAPE and EAPE reservoirs, representing eddy energy generation through baroclinic instability, whereas barotropic energy conversion links the MKE and EKE reservoirs, representing eddy energy generation through barotropic instability. Previous studies have shown that a large fraction of the wind energy input to the ACC is converted into eddy energy through baroclinic energy conversion downstream of major topographic features (Youngs et al. 2017; Wu et al. 2017; Matsuta and Masumoto 2023; Matsuta et al. 2024). This occurs because standing meanders enhance baroclinic growth rates downstream of topography (Thompson and Naveira Garabato 2014; Abernathey and Cessi 2014; Bischoff and Thompson 2014). Matsuta and Masumoto (2023) estimated that more than 70% of baroclinic energy conversion in the ACC is localized downstream of five major topographic obstacles. These studies indicate that flow-topographic interactions control the ACC energy balance. It should be also noted that although many models consistently indicate that baroclinic energy conversion is dominant, the contribution of barotropic energy conversion differs among studies. Some suggest that, when integrated over the entire ACC, its contribution is negligible (Matsuta and Masumoto 2023; Matsuta et al. 2024), whereas others argue that under strengthened westerly winds it can make a non-negligible contribution (Youngs et al. 2017; Wu et al. 2017).

Recent studies have pointed out that baroclinic energy conversion plays an important role in determining the sensitivity of baroclinicity to friction (e.g., Hogg and Blundell 2006; Nadeau and Straub 2012; Nadeau and Ferrari 2015; Marshall et al. 2017; Mak et al. 2017; Yang et al. 2018; Mak et al. 2018; Jouanno and Capet 2020; Klymak et al. 2021; Stewart et al. 2023; Yang et al. 2023; Maddison et al. 2025). Here, *baroclinicity* refers to the meridional slope of the isopycnal surface. These studies showed that as friction increases, baroclinicity and hence circumpolar transport increase. Marshall et al. (2017) explained this counter-intuitive relationship focusing on the eddy momentum transfer and eddy energetics. If transient eddies are responsible for the equilibrium of the ACC, transient eddies should transport surface stress downward to ultimately be removed at the bottom (Johnson and Bryden 1989). They assumed that the eddy interfacial form stress is scaled by eddy energy as $\alpha E$ based on the eddy geometric parametrization (Marshall et al. 2012), where $E$ is the eddy energy and $\alpha$ is a tuning parameter associated with the eddy geometric parameterization. Since the eddy interfacial form stress should balance the wind momentum input if the residual circulation vanishes (Marshall and Radko 2003), the eddy energy is scaled by the wind stress,

$$\alpha E \sim \tau_0, \tag{1}$$

where $\tau_0$ is the wind stress strength. If $\alpha$ can be regarded as a constant, this scaling indicates that eddy energy is constrained by wind stress and not dependent on friction. In order to maintain this momentum balance, the baroclinic energy conversion should sustain the eddy energy against the dissipation $\lambda E$, where $\lambda$ is eddy dissipation rate. Since the baroclinic energy conversion is dependent on the baroclinicity and eddy interfacial form stress, the increase in the friction or eddy energy dissipation rate requires an increase in the baroclinicity,

$$\alpha E s \sim \lambda E,$$

or

$$s \sim \frac{\lambda}{\alpha}, \tag{2}$$

where $s$ is the baroclinicity. Marshall et al. (2017) referred to this mechanism as *frictional control*.

A key feature missing in the frictional control mechanism is flow-topographic interactions. According to Rivière et al. (2004), when the bottom drag coefficient is low, the barotropic mode dominates the mean flow, causing the mean flow to be more strongly influenced by bottom topography. This may result in the increase in the barotropic energy conversion due to an enhanced standing meander downstream of the ridge (Youngs et al. 2017), even though the frictional control mechanism implicitly assumes that EKE is generated only through baroclinic instability. If the barotropic energy conversion is not negligible, the relationship between eddy energy and eddy interfacial form stress may weaken. Indeed, Marshall et al. (2017) discussed that, in the low-drag regime, EKE depends on drag coefficients, indicating a breakdown of (1). Stewart et al. (2023) also suggested that friction influences eddy energy and the strength of the standing meander. However, the frictional dependence of the LEC has received little attention, and the drag regime over which (1) and (2) are applicable remains unclear.

To address this gap, we conduct a series of sensitivity experiments using an eddy-rich reentrant channel model, in which the linear drag coefficient is systematically changed. We evaluate the drag dependence of standing meander and eddy energy in each regime. We also examine how barotropic and baroclinic energy conversions change across different drag regimes using the LEC. By combining these analyses, we propose a generalized frictional control relation that extends the formulation of previous studies.

This study is organized as follows. Section 2 describes the model configurations. In Section 3, we investigate the dependency of eddy energy and LEC on the drag coefficient through numerical experiments. Section 4 proposes a generalized frictional control. Section 5 provides a summary of this study.

## 2. Methods

*2.1 Model configuration*

We use MITgcm (Marshall et al. 1997) in the Boussinesq and hydrostatic approximations on the $\beta$ plane. The domain is 4000-km long (zonal, $x$-direction), 2000-km wide (meridional, $y$-direction), and 3881-m deep (vertical, $z$-direction). A free-slip boundary condition is imposed at the northern and southern boundaries and a periodic boundary condition is in the zonal direction. There is no heat transport across the southern boundary. We used a Cartesian grid with a horizontal resolution of $10 \times 10$ km to resolve mesoscale eddies. There are 29 vertical levels increasing in thickness from 8.5 m at the ocean surface to 248 m at depth. The Coriolis parameter is $f = f_0 + \beta y$, with $f_0 = -10^{-4}\ \mathrm{s}^{-1}$ and $\beta = 10^{-11}\ \mathrm{m}^{-1}\mathrm{s}^{-1}$. We used a reference density $\rho_0$ of $1035\ \mathrm{kg\ m}^{-3}$. The density is prescribed by a temperature-only linear equation of state with a thermal expansion coefficient of $2.0 \times 10^{-4}{}^{\circ}\mathrm{C}^{-1}$. The subgrid horizontal and vertical eddy viscosity are $100\ \mathrm{m}^2\ \mathrm{s}^{-1}$ and $3.0 \times 10^{-4}\ \mathrm{m}^2\ \mathrm{s}^{-1}$, respectively, while the subgrid horizontal and vertical eddy diffusivities are zero and $5.0 \times 10^{-6}\ \mathrm{m}^2\ \mathrm{s}^{-1}$. To represent a surface mixed layer, we employed the K-profile parameterization (KPP) mixing scheme (Large et al. 1994). The bottom topography consists of a Gaussian ridge centered at $x_0 = 1000$ km:

$$\eta_b(x, y) = h_0 \exp\left[-\frac{(x - x_0)^2}{\sigma_x^2}\right], \tag{3}$$

where $h_0 = 1500$ m and $\sigma_x = 300$ km. Hence, the ocean thickness is given as $h = H_0 - \eta_b$, where $H_0 = 3881$ m. The wind stress is expressed as:

$$\tau_w(y) = \tau_0 \sin\left(\pi \frac{y}{L_y}\right), \quad 0 \le y \le L_y, \tag{4}$$

where $\tau_0 = 0.2\ \mathrm{N\ m}^{-2}$ and $L_y = 2000$ km is the meridional extent of the domain. At the northern boundary, the temperature is relaxed to an exponential profile ranging from 8°C at the surface to 0°C at the bottom:

$$\theta(z) = \frac{8\left(e^{\frac{z}{h_{sc}}} - e^{-\frac{H_0}{h_{sc}}}\right)}{\left(1 - e^{-\frac{H_0}{h_{sc}}}\right)}, \tag{5}$$

where $\theta$ is the potential temperature and $h_{sc} = 1000$ m is the scale height. The initial stratification is also based on this profile.

The bottom drag is represented using a linear function of the bottom velocity. A central aim of this work is to investigate how eddy energetics depend on the bottom linear drag coefficient:

$$r = \{10^{-2}, 7.5 \times 10^{-3}, 5.0 \times 10^{-3}, 2.5 \times 10^{-3}, 10^{-3}, 7.5 \times 10^{-4}, 5.0 \times 10^{-4}, 2.5 \times 10^{-4}, 10^{-4}, 7.5 \times 10^{-5}, 5.0 \times 10^{-5}, 10^{-5}, 10^{-7}\}[\mathrm{m\ s^{-1}}].$$

Although dissipation involves complex processes (e.g., lee waves, inverse cascades, and spontaneous emission), previous studies suggest that increases in form drag due to unresolved mesoscale topographic roughness can be effectively represented by an increased linear drag coefficient (Klymak 2018; Klymak et al. 2021). Therefore, different values of the linear drag coefficient can be interpreted as representing different levels of topographic roughness.

Because a stratified channel model reaches mechanical equilibrium within approximately 40 years (Ward and Hogg 2011), all simulations were integrated for 50 years, with the final five years utilized for calculations. For representative cases, simulations were integrated for 100 years, and we confirmed that our results were not dependent on the averaging periods. It should be noted that reaching a thermal equilibrium state would require integrations of roughly 1000 years (Munday et al. 2015), but this is not computationally feasible.

### *2.2 Energy conversion between eddy and mean flow*

We calculate the energy cycle of reentrant channels by considering Lorenz energy cycle (LEC) (Lorenz 1955) linking mean kinetic energy (MKE), $K_M$, eddy kinetic energy (EKE), $K_E$, mean available potential energy (MAPE), $P_M$, and eddy available potential energy (EAPE), $P_E$ (Figure 1). The derivation of this diagram follows standard textbooks (e.g., Olbers et al. 2012).

At any given location $\mathbf{x} = (x, y, z)$ we define:

$$K_M(x, y, z) = \frac{1}{2}\rho_0\left(\overline{u}^2 + \overline{v}^2\right), \tag{6}$$

$$K_E(x, y, z) = \frac{1}{2}\rho_0\left(\overline{u'^2} + \overline{v'^2}\right), \tag{7}$$

$$P_M(x, y, z) = -\frac{1}{2}\frac{g}{\partial_z \rho_{bg}(z)}\,\overline{\rho^*}^2, \tag{8}$$

and

$$P_E(x, y, z) = -\frac{1}{2}\frac{g}{\partial_z \rho_{bg}(z)}\,\overline{\rho'^2}, \tag{9}$$

where $\mathbf{u} = (u, v, w)$ is the velocity field, $g = 9.81\ \mathrm{m\ s^{-2}}$ is the gravitational acceleration, the overbar denotes a time average (in our simulations, taken over the final 5-year period), and the prime denotes a transient eddy component defined as the deviation from the time average. References to *eddy* denote a transient eddy hereafter. The background density $\rho_{bg}$ is defined as horizontally averaged mean density. The deviation of density from the background state is defined as

$$\rho^*(t, x, y, z) = \rho(t, x, y, z) - \rho_{bg}(z). \tag{10}$$

With this definition,

$$\rho' = \rho - \overline{\rho} = \left(\rho^* + \rho_{bg}\right) - \overline{\rho^* + \rho_{bg}} = \rho^* - \overline{\rho^*}. \tag{11}$$

In our simulations, the energy input is described as:

$$I_W = \iint \boldsymbol{\tau}_w \cdot \overline{\mathbf{u}}_h^{(0)}\, dxdy\,, \tag{12}$$

where $\boldsymbol{\tau}_w$ is the surface wind stress and $\mathbf{u}_h^{(0)}$ is the surface horizontal flow. The integration is performed over the channel. Note that although the westerly wind forcing is identical across the numerical experiments, the energy input depends on the modeled $\mathbf{u}_h^{(0)}$, which differs depending on the drag coefficient. The objective of this study is not to examine how the relative proportions of energy conversion pathways change under a fixed energy input; rather, it is to investigate how the entire energy diagram—including the energy input itself—changes when the same westerly wind forcing is applied.

The energy supplied by the wind is redistributed among the four energy reservoirs through energy conversion terms, as shown in Figure 1. The conversion between MKE and MAPE is:

$$C(K_M, P_M) = \iiint g\overline{\rho^*}\,\overline{w}\,dxdydz = \iiint \overline{\mathbf{u}}_h \cdot \overline{\nabla_h p}\ dxdydz$$
$$= \iiint \overline{\mathbf{u}_{h,a}} \cdot \overline{\nabla_h p}\ dxdydz, \tag{13}$$

where $p$ is pressure, $\mathbf{u}_h = (u, v)$ is the horizontal velocity field, $\mathbf{u}_{h,a}$ is the ageostrophic component, and $\nabla_h$ is the horizontal gradient operator. A positive sign for $C(K_M, P_M)$ indicates energy draining to MAPE from MKE.

The conversion from MAPE to EAPE is

$$C(P_M, P_E) = \iiint \frac{g}{\partial_z \rho_{bg}(z)} \overline{\rho' \mathbf{u}_h'} \cdot \nabla_h \overline{\rho^*}\, dxdydz\,, \tag{14}$$

which is the baroclinic conversion rate (BCR), indicating, when positive, that eddies are extracting potential energy from the mean stratification. Because a mixed layer exists in roughly the upper 400 m, the vertical density gradient becomes extremely small there, which leads to unrealistically large values of BCR. To avoid this problem, the background density gradient in the upper 450 m is estimated as the density gradient between the ocean surface and 450 m depth. Similarly, the conversion from EAPE to EKE is:

$$C(P_E, K_E) = -\iiint g\overline{\rho' w'}\, dxdydz\,, \tag{15}$$

which is the vertical eddy density flux (VEDF). A positive value of VEDF corresponds to the EKE generation by releasing EAPE. We refer to the energy pathway of $K_M \rightarrow P_M \rightarrow P_E \rightarrow K_E$ as *baroclinic energy pathway*.

The barotropic energy conversion rate (BTR), $C(K_M, K_E)$ is defined as:

$$C(K_M, K_E) = \rho_0 \iiint \left(\overline{u}\nabla \cdot \overline{u'\mathbf{u'}} + \overline{v}\nabla \cdot \overline{v'\mathbf{u'}}\right) dxdydz$$

$$= -\rho_0 \iiint \left(\overline{u'\mathbf{u'}} \cdot \nabla\overline{u} + \overline{v'\mathbf{u'}} \cdot \nabla\overline{v}\right) dxdydz\,, \quad (16)$$

where $\nabla$ is the three-dimensional gradient operator. A positive sign indicates EKE generation by barotropic instability (Pedlosky 1987). We refer to the energy pathway of $K_M \rightarrow K_E$ as *barotropic energy pathway*.

The energy inputted to the system is finally balanced by MKE and EKE dissipation by viscosity and bottom friction, indicated by $D(K_M)$ and $D(K_E)$, respectively. In addition, the northern sponge layer and diffusivity act as both source and sink for the APE, which are merged into $D(P_M)$ and $D(P_E)$ in the energy diagram.

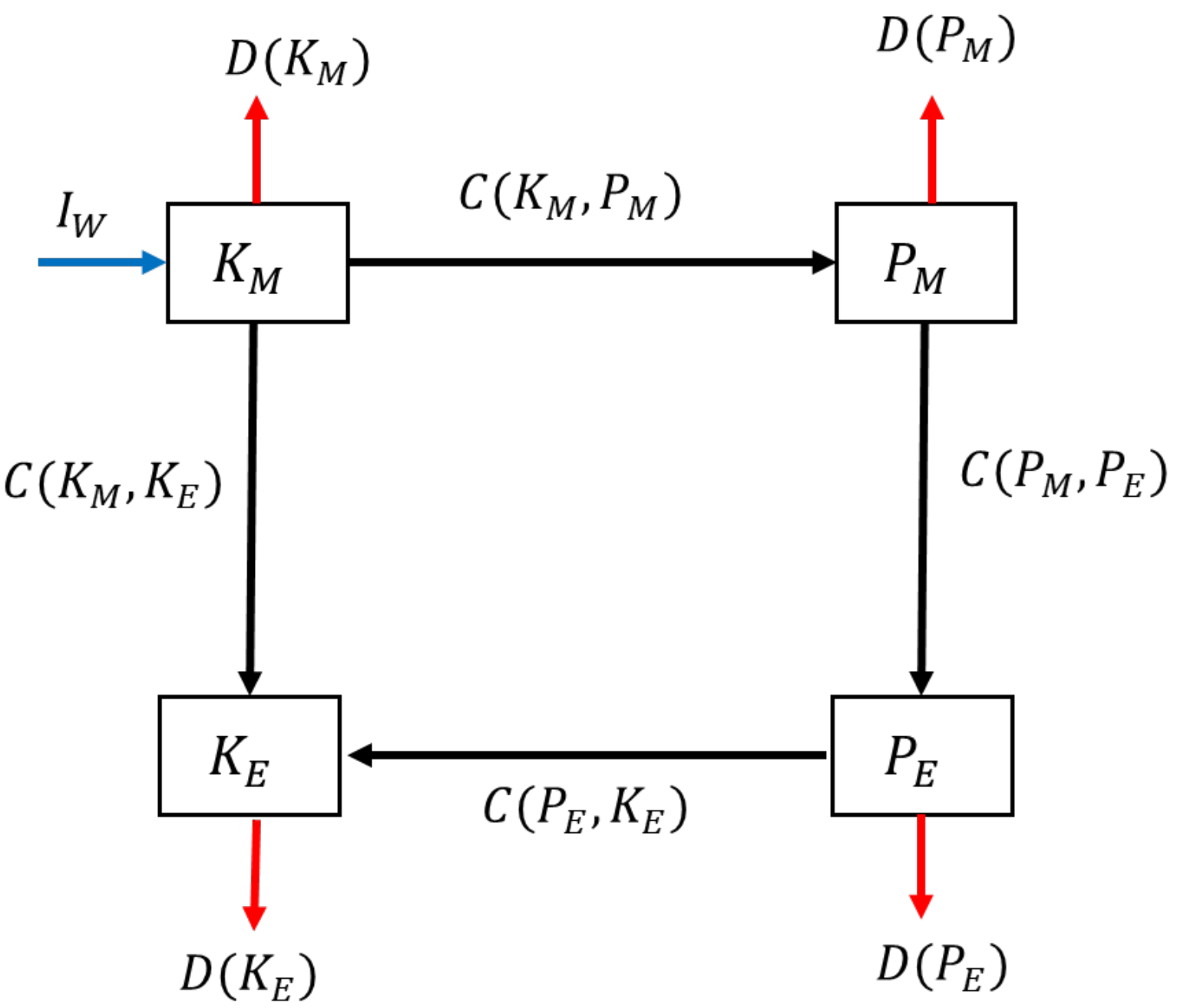

Figure 1. Four-box energy diagram among $K_M$, $K_E$, $P_M$, and $P_E$. Each reservoir is connected by energy conversion functions indicated by $C(*,*)$. Arrow directions correspond to the sign convention of each conversion term. Energy generation and dissipation by subgrid processes in each reservoir is denoted by $D(*)$, indicated by red arrows. In our simulation, the reentrant channels are driven by westerlies ($I_W$ in the diagram, indicated by a blue arrow).

## 3. Results

*3.1 Basic description of mean flow in each drag regime*

We first present the dependence of horizontal circulation on the drag coefficient. The barotropic streamfunction is defined as:

$$\psi(x,y) = \int_0^y \int_{-h}^0 \overline{u}\, dzdy\,. \tag{17}$$

We compare the outputs of $r = 10^{-2}$ m s$^{-1}$ , (termed "HIGH"), $r = 5 \times 10^{-4}$ m s$^{-1}$ (termed "MEDIUM"), and $r = 10^{-5}$ m s$^{-1}$ (termed "LOW") as representative cases of high-drag, medium-drag, and low-drag conditions, respectively. Figure 2 demonstrates that large EKE values are localized in topographic lees, suggesting that eddy activities are constrained by local dynamics as suggested in previous studies (Thompson and Naveira Garabato 2014; Abernathey and Cessi 2014), regardless of the bottom drag coefficient. By contrast, the EKE magnitude appears to depend on friction. We return to this point in the next section.

A key difference in horizontal circulation among the three cases is seen in the strength of wind-driven gyre and standing meander. Figure 2 shows that the curvature of standing meanders increases as drag coefficient reduces. In addition, wind-driven gyre circulations are more evident with decreasing drag. According to Nadeau and Ferrari (2015), weaker friction increases the sensitivity of mean flow to bottom topography, allowing a ridge to act as an effective western boundary and the mean flow to be governed by Sverdrup dynamics. The differences in sensitivity to bottom topography may be explained by the barotropization of the mean flow. Figure 3 shows the fraction of MKE of the barotropic mode relative to total MKE, integrated over the domain, as a function of drag coefficient. The barotropic mode intensifies as the bottom drag coefficient decreases for $r \geq 10^{-5}$ m s$^{-1}$, and the fraction reaches 85 % at $r = 10^{-5}$ m s$^{-1}$, indicating that the mean flow is approximately barotropic under low-drag conditions. This occurs because the bottom current experiences minor damping under low drag, and the opposite is true for high drag (Rivière et al. 2004; LaCasce 2017). It should be noted that the barotropic mode becomes insensitive to the drag coefficient for $r \leq 10^{-5}$ m s$^{-1}$ because the timescale of spindown, $(r/H_0)^{-1}$, is greater than 12 years. Hence, the bottom friction can be neglected in this regime, and the viscosity is responsible for the dissipation.

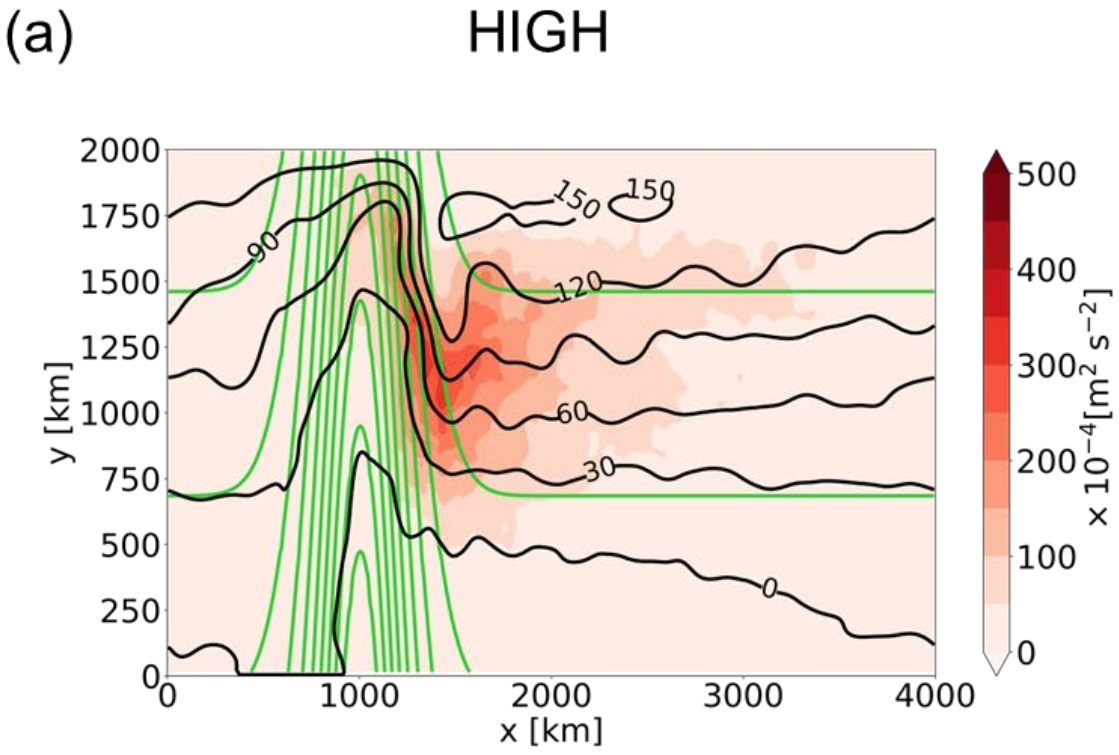

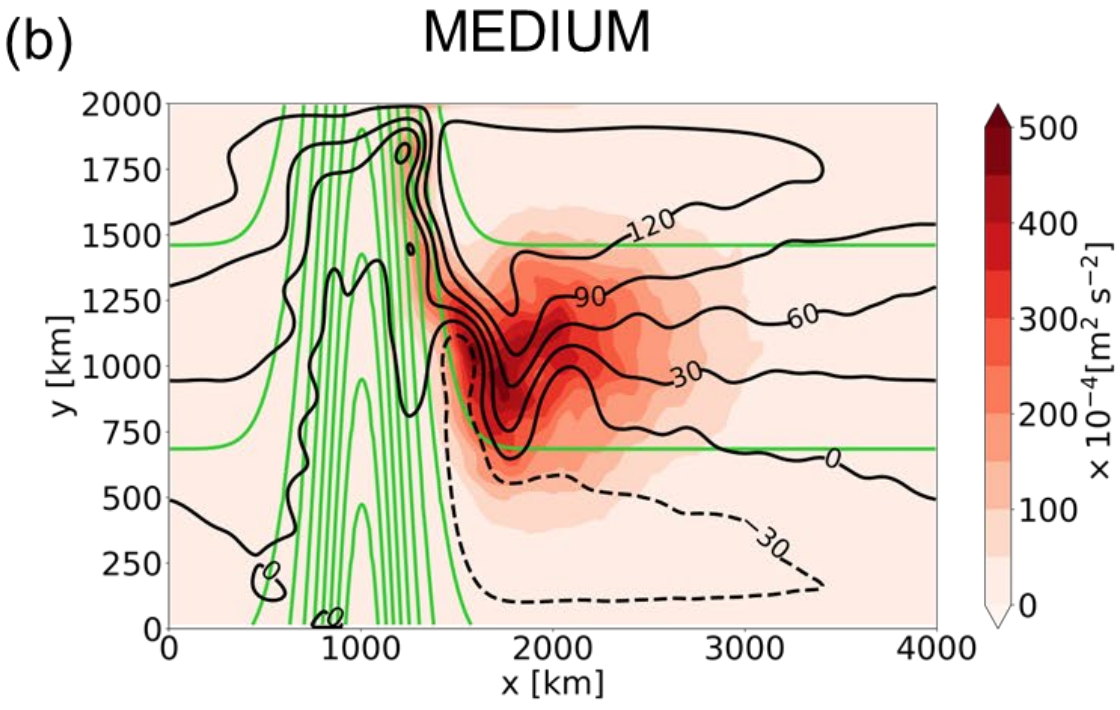

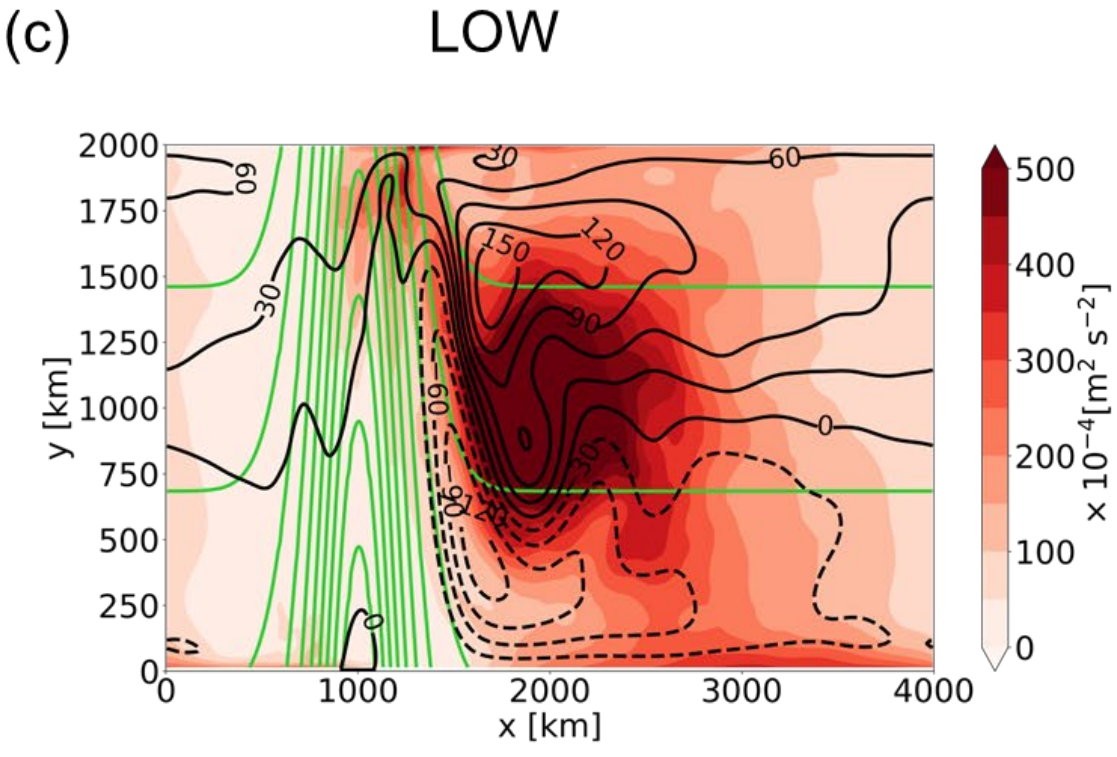

Figure 2. Horizontal distribution of vertically averaged EKE per density (background colors) and barotropic streamfunction contours (black lines, [Sv]) for (a) HIGH, (b) MEDIUM, and (c) LOW. Green lines are geostrophic contours with a contour interval of $2.0 \times 10^{-9}$ m$^{-1}$ s$^{-1}$. Here, the geostrophic contour is defined as $-(f_0 + \beta y)/h$.

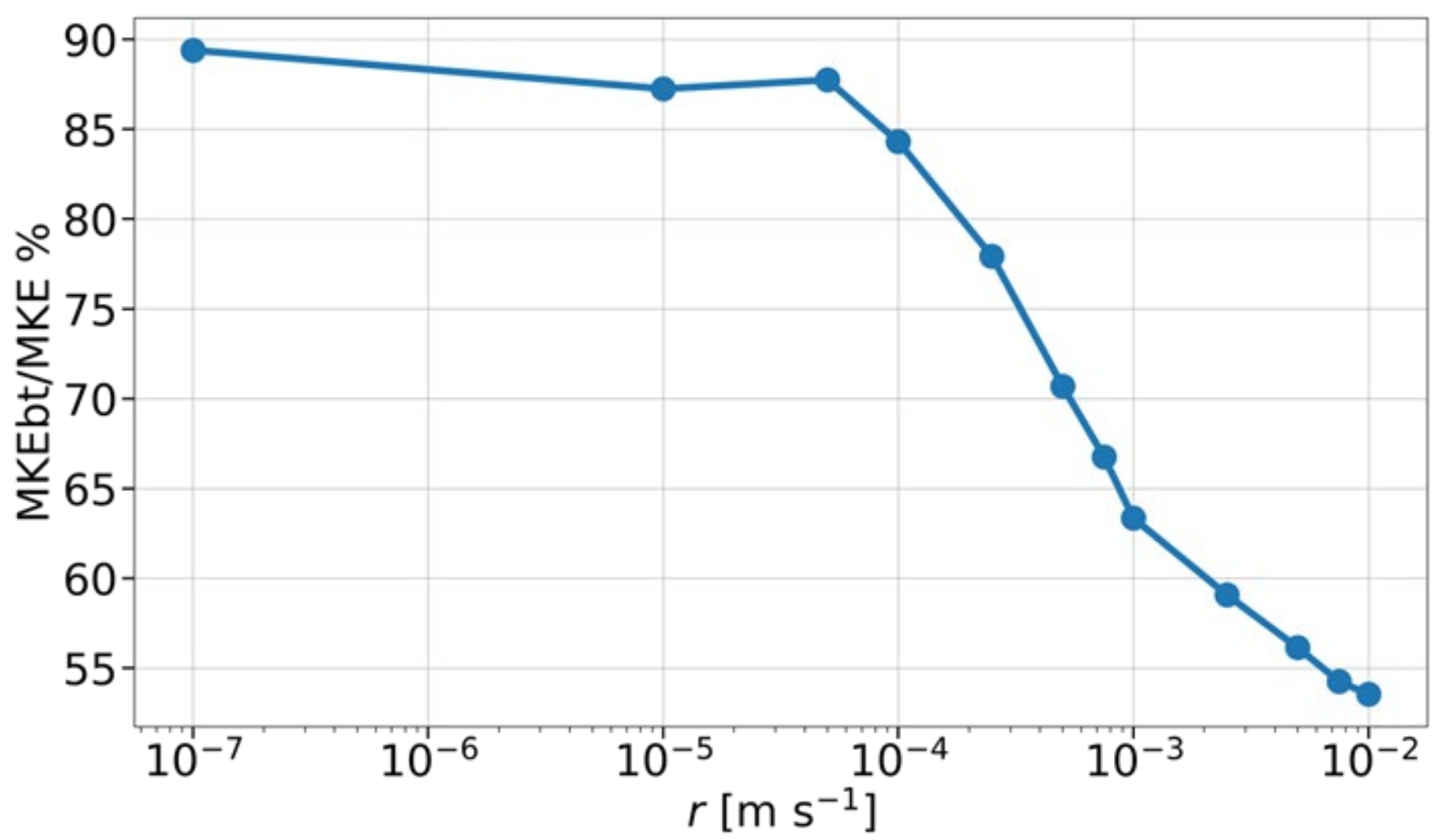

Figure 3. Fraction of the barotropic mode of MKE relative to the total MKE.

*3.2 Dependency of eddy energy*

Figure 4 provides a quantitative illustration of the dependence of eddy energy on friction. The EKE values reduce with increasing drag coefficient for $r \leq 10^{-3}$ m s$^{-1}$, while it is less sensitive in the high-drag regime. This result is consistent with Figure 2 in Marshall et al. (2017). In contrast to EKE, the EAPE values increase with drag coefficient. As a result, the total eddy energy, the sum of EKE and EAPE decreases with the drag coefficient in the regime of $r \leq 10^{-3}$ m s$^{-1}$, while it increases in the high-drag regime. The friction-induced increase in total eddy energy in the high-drag regime has also been reported in Mak et al. (2018). These results suggest that eddy energy strongly depends on friction except around $r = 10^{-3}$ m s$^{-1}$.

The disagreement between (1) and numerical results (Figure 4) likely stems from the assumption that the eddy geometric parameter is constant and independent of friction. Previous studies have suggested that adopting a spatially uniform $\alpha$ over the entire channel is not particularly problematic when the drag coefficient is fixed (Marshall et al. 2012; Poulsen et al. 2019); by contrast, our result suggests that $\alpha$ may change substantially when the drag coefficient is varied as suggested by Mak et al. (2017).

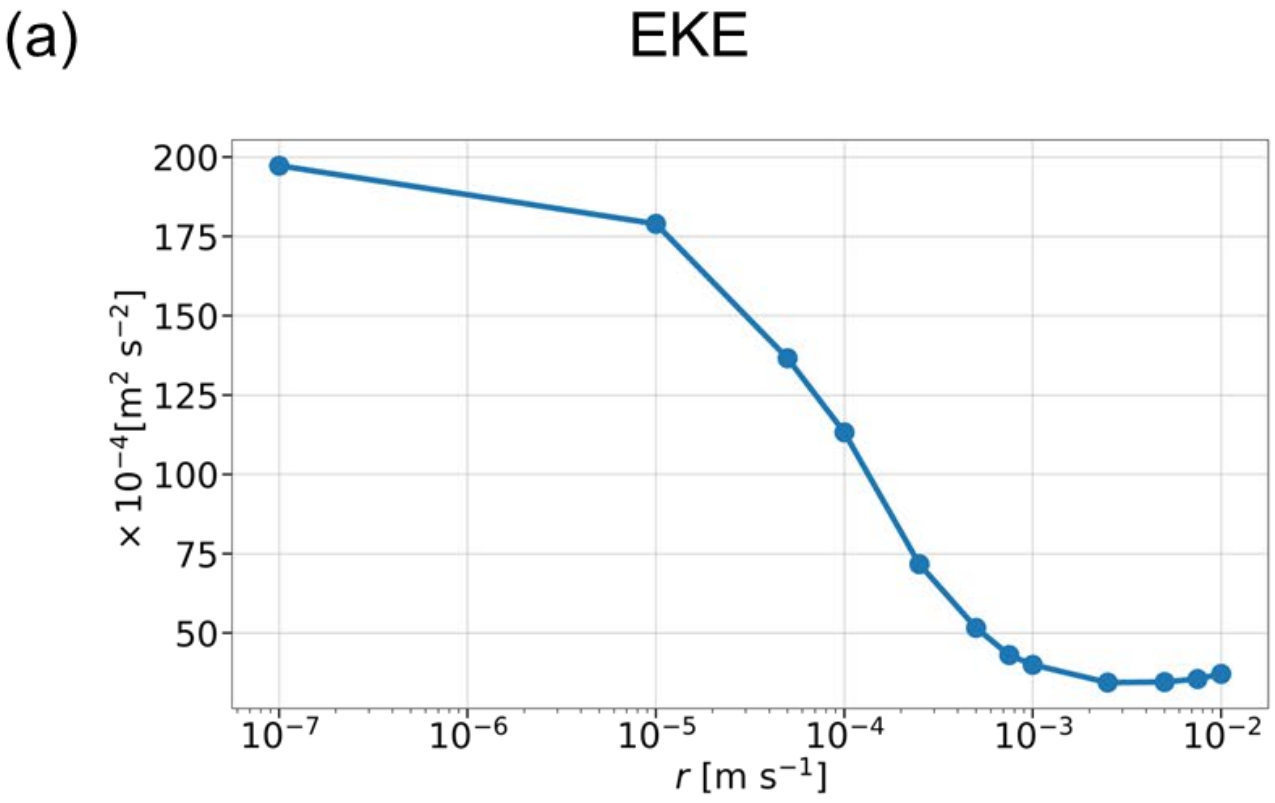

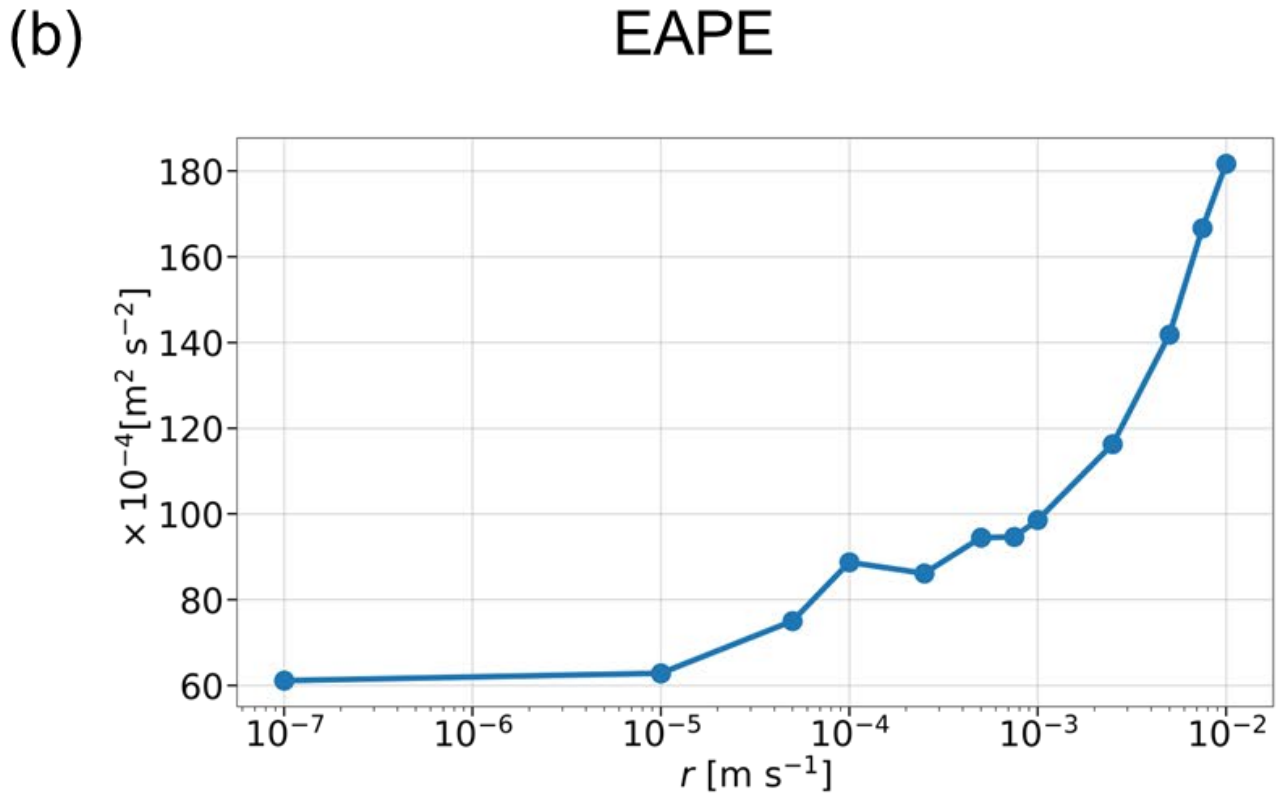

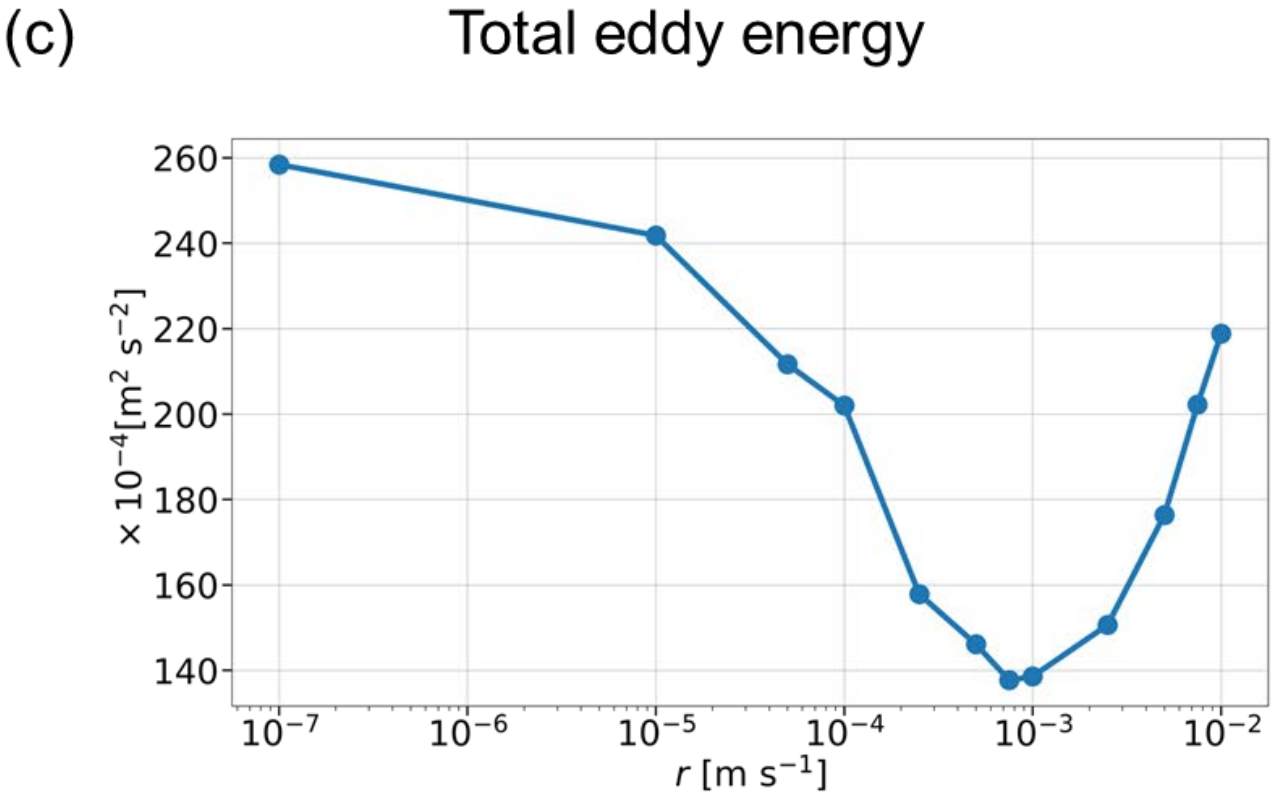

Figure 4 Blue curves indicate domain-averaged (a) EKE, (b) EAPE, and (c) total eddy energy per density as functions of drag coefficients.

*3.3 Dependence of BCR and BTR on the linear drag coefficient*

We examine the horizontal distribution of VEDF and BTR (Figure 5). We analyze VEDF here rather than BCR because it is unaffected by the rotational component of the eddy flux (Marshall and Shutts 1981) and thus clearly reflects the net distribution of baroclinic instability. The choice between these two quantities does not change the domain-integrated energy budget analysis and hence the qualitative conclusion. In the HIGH case, VEDF exhibits large positive values downstream of the topography because the standing meander locally intensifies baroclinicity (Thompson and Naveira Garabato 2014; Bischoff and Thompson 2014). For the MEDIUM and LOW cases, the spatial distribution exhibits similar characteristics. However, the amplitude of VEDF decreases with decreasing drag coefficient. In particular, VEDF is confined along the western boundary in the LOW case. The lower panel of Figure 5 demonstrates that the BTR is also localized downstream of topography. In HIGH, the BTR exhibits a patchy structure with both positive and negative values. A similar distribution is found in the ACC of the high-resolution models (Wu et al. 2017; Matsuta and Masumoto 2023). As friction decreases and the standing meander becomes larger, the area of positive BTR widens. Although negative regions still exist in the low-drag regime, as shown below, the BTR becomes positive when integrated over the domain.

To demonstrate the impact of drag coefficient quantitatively, we examine the LEC for HIGH, MEDIUM and LOW. Figure 6 shows that the fate of wind energy varied with the drag coefficient. Under high-drag conditions (Figure 6(a)), nearly 65% of the wind energy is converted to the MAPE. The APE gain is converted into EKE via BCR (86 GW) and VEDF (71 GW). By contrast, the barotropic conversion is negative, implying a transfer of energy from EKE to MKE. This behavior is consistent with the barotropic governor mechanism (James and Gray 1986; Solodoch et al. 2016; Youngs et al. 2017). The barotropic governor refers to the process by which eddies generated through baroclinic conversion produce upgradient momentum fluxes, leading to the extraction of EKE by the horizontal shear of the mean flow. According to Youngs et al. (2017), eddies tilt in the upgradient direction downstream of topography in a stratified reentrant channel model, which is reflected in the negative BTR. Consequently, in the high-drag regime, eddy energy is primarily governed by the baroclinic pathway, with the barotropic pathway providing a weak damping effect.

By contrast, under low-drag conditions (Figure 6(c)), the barotropic route substantially contributes to the EKE generation. Although some negative BTR regions are present, particularly at the northern edge of the southern gyre, positive BTR is dominant (Figure 5(f)), indicating that enhanced standing meanders facilitate eddy generation via barotropic instability. Negative BTR areas (Figure 5(f)) can arise even if barotropic instability is substantial owing to several processes, including the barotropic governor mechanism, the phase of the standing meander (Takaya and Nakamura 2001; Danielson et al. 2006), and nonlocal eddy–mean flow interactions (Murakami 2011; Chen et al. 2014; Matsuta and Masumoto 2021). Accordingly, we refrain from attributing the negative BTR identified in Figure 5(f) to any single mechanism. In contrast to the enhanced BTR, baroclinic energy pathway is weaker than in the high-drag case: the reduced conversion from MKE to MAPE results in smaller BCR and VEDF. Under medium-drag conditions (Figure 6(b)), the partitioning of energy inputs exhibited characteristics between these two regimes. It should also be noted that the MKE sink shows little change among experiments. Under low-drag conditions, the MKE sink at the ocean floor is naturally reduced; however, this is offset by increased viscous dissipation in the ocean interior due to an enhanced standing meander.

To quantify the dependency of energy conversion rates on the bottom drag, the energy conversion rates are plotted as functions of drag coefficients in Figure 7(a). If the bottom drag is smaller than $10^{-4}$ m s$^{-1}$, the contribution of BTR is substantial. The BTR values decreases to zero between $r = 10^{-4}$ m s$^{-1}$ and $r = 10^{-3}$ m s$^{-1}$ and takes small negative values in the high-drag regime. Conversely, $C(K_M, P_M)$, BCR and VEDF increase in tandem with increasing drag coefficient, suggesting that increased bottom drag intensifies the baroclinic energy pathway. The mechanism by which MAPE gain increases with friction is discussed in Appendix. The compensating relationship between barotropic and baroclinic pathways causes the BCR fraction in the eddy energy generation, $\mathrm{BCR}/(\mathrm{BCR} + \mathrm{BTR})$ to vary with the drag coefficient. Figure 7(b) demonstrates that the BCR fraction in the EKE gain is 75 % under the low-drag conditions below $r = 10^{-4}$ m s$^{-1}$. It increases sharply to 100 % between $r = 10^{-4}$ m s$^{-1}$ and $r = 10^{-3}$ m s$^{-1}$. Therefore, in the low- and medium-drag regimes, eddy energy is partly generated by barotropic instability. As a result, eddy energy does not serve as an indicator of eddy interfacial form stress in these regimes, which may reflect in the breakdown of (1).

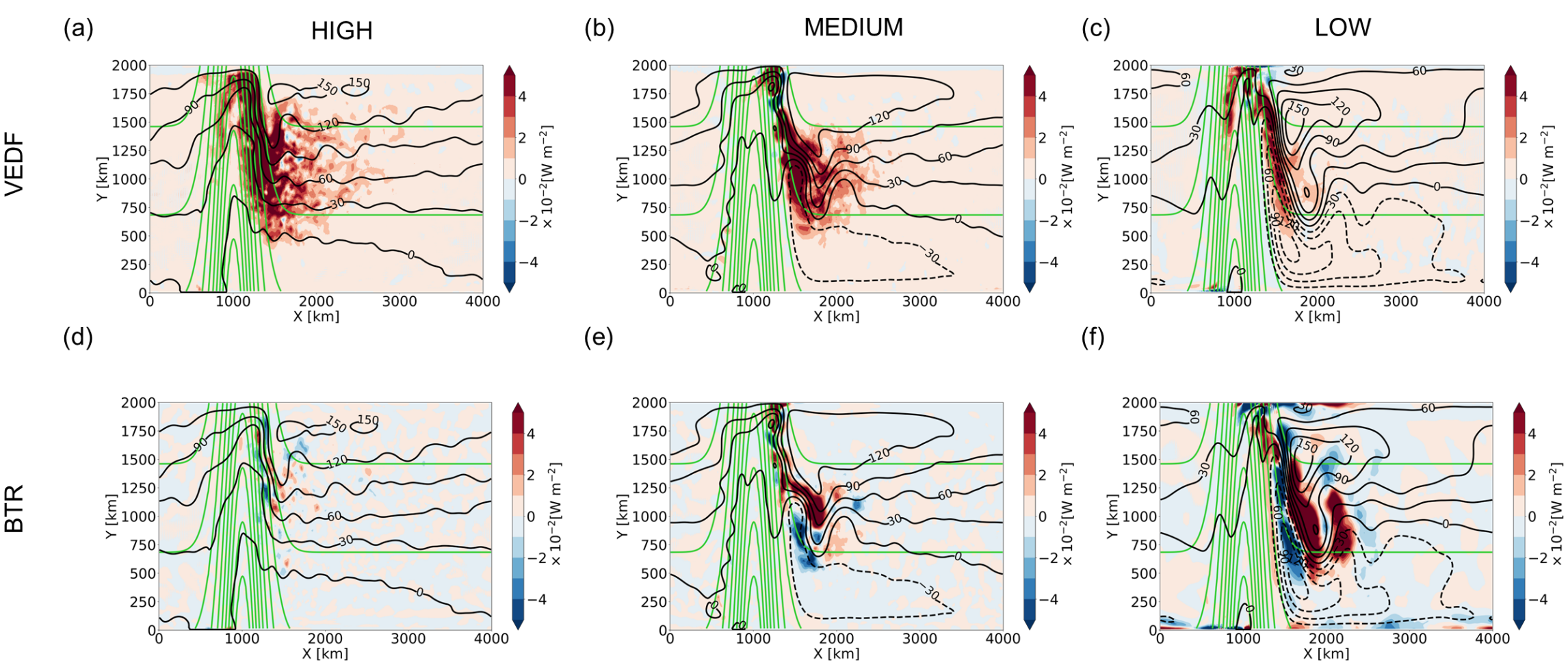

Figure 5 Horizontal distributions of VEDF for (a) HIGH, (b) MEDIUM, and (c) LOW. Green and black contours are the same as those in Figure 2. Lower panels are the same as the upper panel but for BTR.

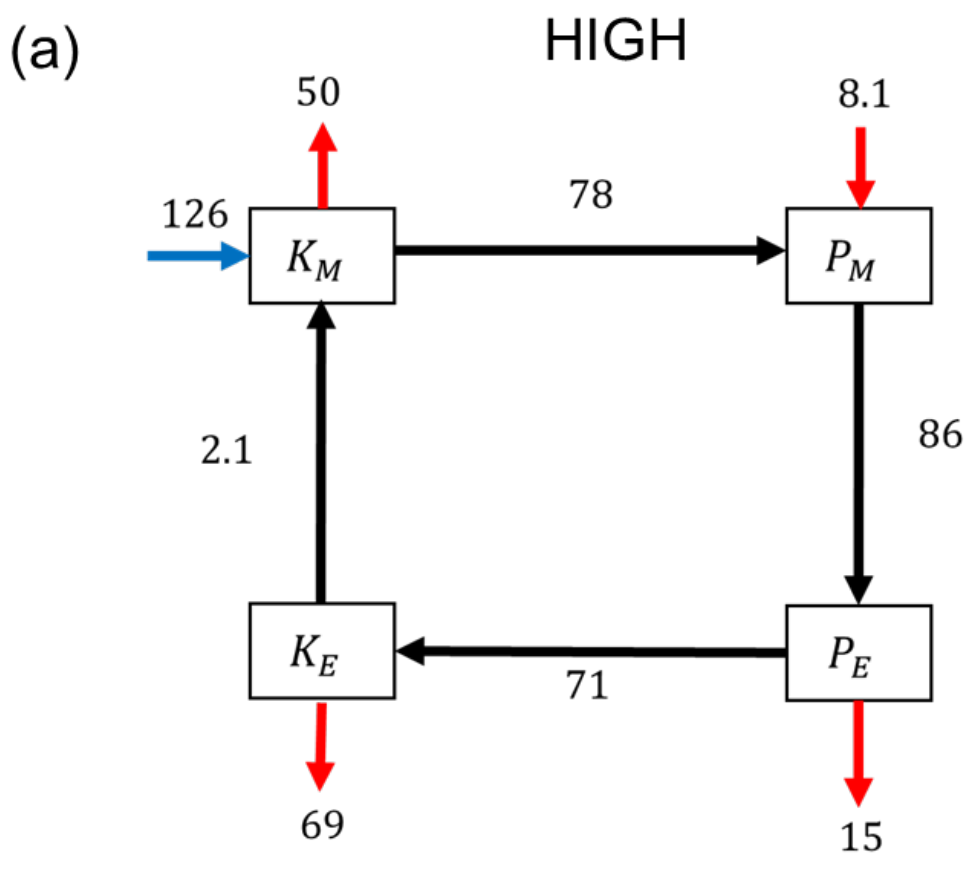

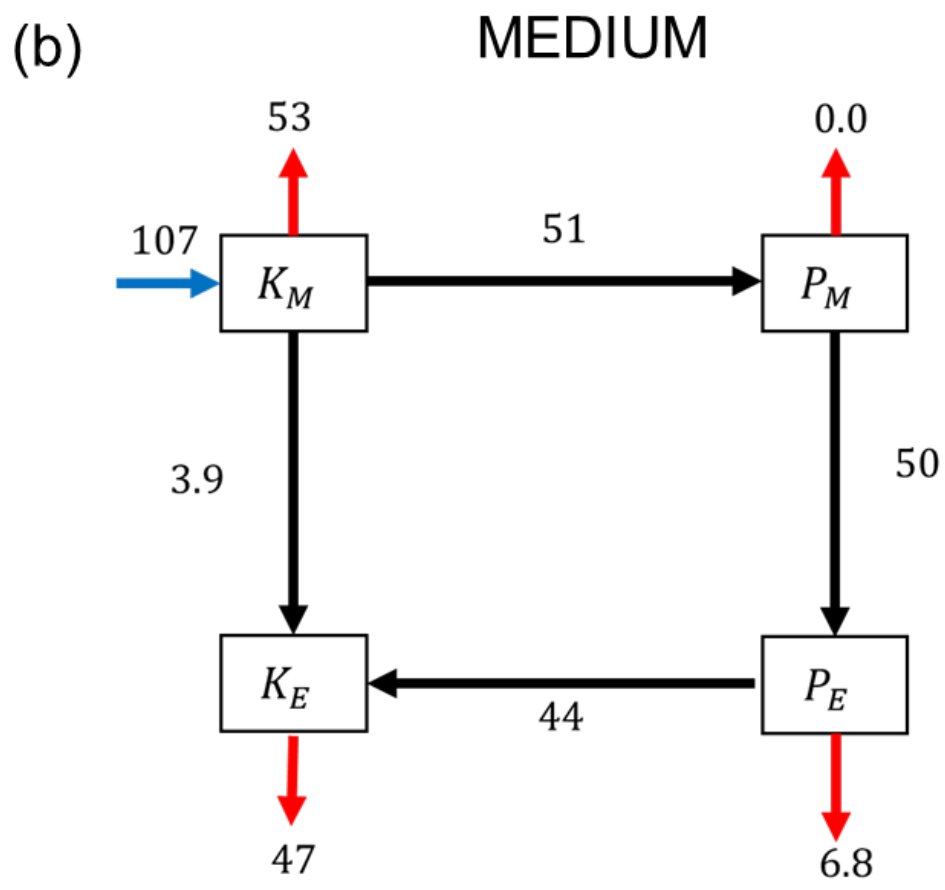

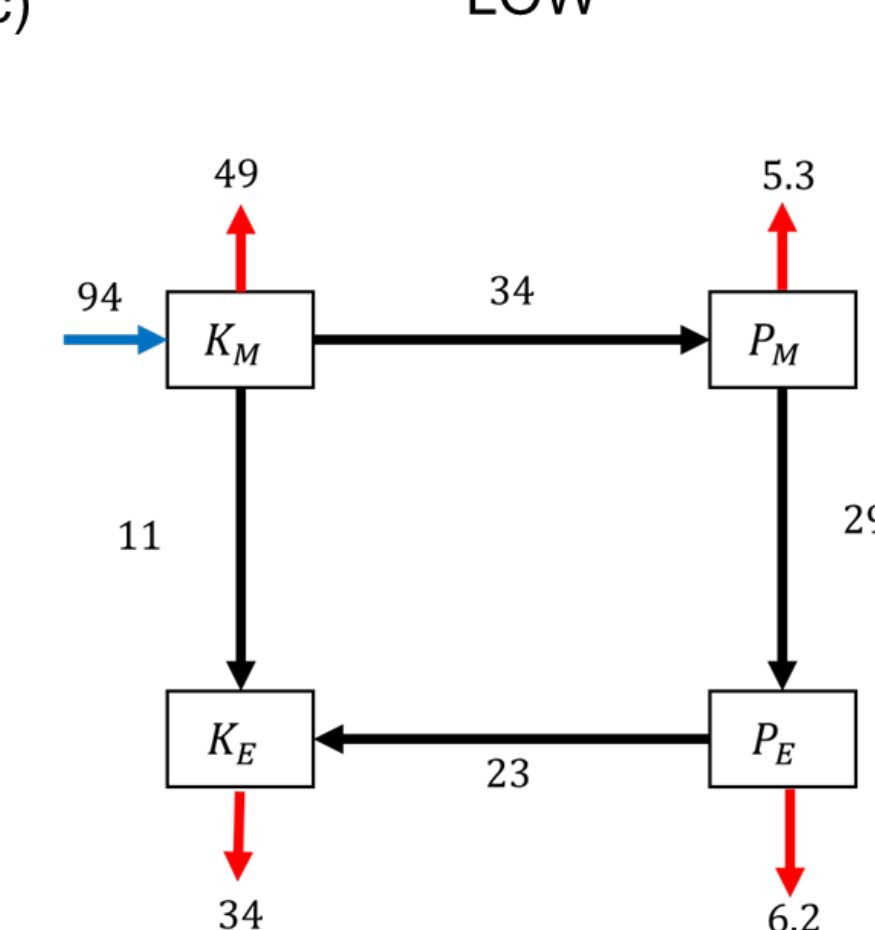

Figure 6. LEC in the same style as Figure 1 calculated for (a) HIGH, (b) MEDIUM, (c) and LOW. The arrow directions indicate the direction of energy conversion.

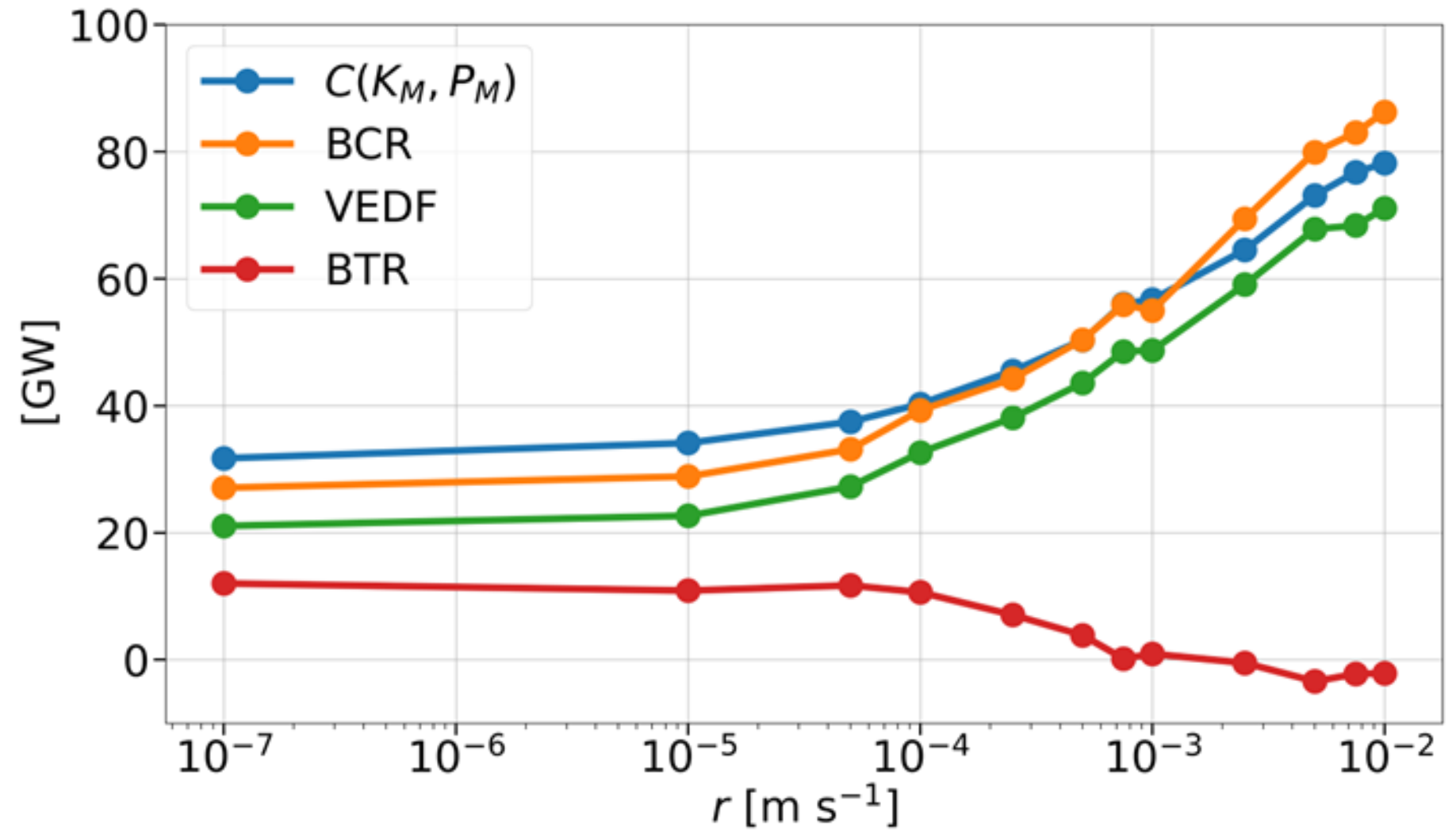

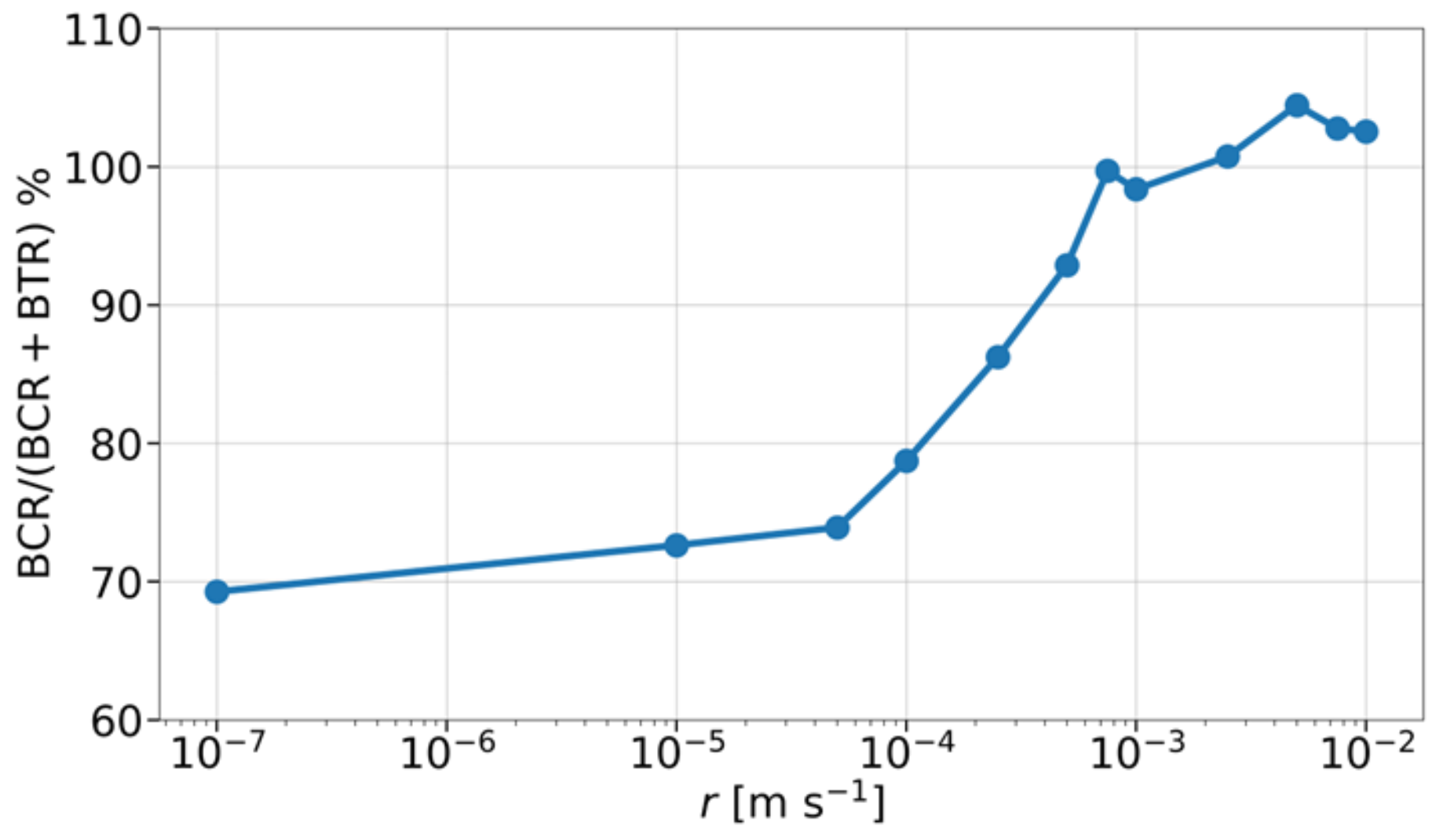

Figure 7. (a) Curves of $C(K_M, P_M)$ (blue line), BCR (orange line), VEDF (green line), and BTR (red line) as functions of drag coefficients. (b) Curves of the fraction of BCR to the sum of BCR and BTR as a function of drag coefficients.

## 4. Discussion: Generalization of frictional control

Figure 8 illustrates how baroclinicity varies with the drag coefficient. Here, the baroclinicity is defined as the difference in the depth of the 1°C isotherm between $y = 500$ km and $y = 1500$ km in the zonally-averaged mean temperature. The baroclinicity is found to increase with friction even outside the high-drag regime. In the framework of frictional control proposed in previous studies, both (1) and (2) should hold. However, Figure 8 suggests that (2) solely holds even if (1) is not valid (Figure 4).

To reconcile this gap, we generalize the frictional control framework based on the LEC without any specific eddy parametrization. The total eddy energy balance is denoted as

$$\mathrm{BCR} + \mathrm{BTR} = D(K_E) + D(P_E), \tag{18}$$

where $D(K_E)$ and $D(P_E)$ are the EKE and EAPE dissipation, respectively. BCR approximately balances $C(K_M, P_M)$ (Figure 7(a)), which, as shown in the Appendix in detail, is primarily due to the work done by the pressure gradient associated with the sea surface elevation acting on the Ekman flow, $C^{(E)}(K_M, P_M)$, with a minor contribution from the pressure gradient acting on the ageostrophic component in the ocean interior. From a simple calculation, $C^{(E)}(K_M, P_M)$ can be expressed in terms of the surface geostrophic velocity,

$$C^{(E)}(K_M, P_M) = \iint \rho_0 g\, \mathbf{U}_{\mathrm{EK}} \cdot \nabla_h \overline{\eta}\, dxdy = \iint \boldsymbol{\tau}_w \cdot \overline{\mathbf{u}}_g^{(0)}\, dxdy\,, \tag{19}$$

where $\mathbf{U}_{\mathrm{EK}}$ is the Ekman transport, $\eta$ is the sea surface elevation, $\boldsymbol{\tau}_w$ is the steady wind stress vector, and $\mathbf{u}_g^{(0)} = \left(u_g^{(0)}, v_g^{(0)}\right)$ is the surface geostrophic current (Roquet et al. 2011). The magnitude of $\overline{\mathbf{u}}_g^{(0)}$ is primarily determined by the baroclinic mode, so that the first term of the left-hand side of (18) is scaled as

$$\mathrm{BCR} \approx \iint \boldsymbol{\tau}_w \cdot \overline{\mathbf{u}}_g^{(0)}\, dxdy \sim \tau_w s, \tag{20}$$

where $s$ is baroclinicity and $\tau_w$ is westerly magnitude. If BTR is smaller than BCR, (18) and (20) gives

$$s \sim \frac{D(K_E) + D(P_E)}{\tau_w}, \tag{21}$$

which suggests that an increase in eddy energy dissipation leads to an enhancement of baroclinicity under a fixed wind stress. Physically, this implies that, as proposed by Marshall et al. (2017), in order to maintain eddy activities against increased eddy energy dissipation, the increased baroclinicity is required to enhance BCR. Therefore, $(21)$ is a generalization of $(2)$.

To validate the scaling of $(21)$, the baroclinicity is compared with $D(K_E) + D(P_E)$ from our experiments. The black dashed curve in Figure 8 shows a function that is proportional to $D(K_E) + D(P_E)$ with a constant offset. It captures well the tendency for baroclinicity to increase with increasing friction in all regimes. In particular, in the range of $r \geq 5.0 \times 10^{-4}$ m s$^{-1}$, where the BCR is dominant in the eddy energy generation (Figure 7), the changes in eddy energy dissipation and baroclinicity agree very well. In the low-drag regime between $r = 10^{-5}$ m s$^{-1}$ and $r = 10^{-4}$ m s$^{-1}$, where BTR cannot be neglected, eddy energy dissipation slightly overestimates the change in baroclinicity.

Before closing this section, it is noted that the generalized frictional control $(21)$ includes the implication for *eddy saturation*. Here, eddy saturation indicates the insensitivity of baroclinicity on wind stress. If the eddy-saturated state is achieved, $(21)$ requires that eddy energy dissipation scales with wind stress. In other words, the condition of $D(K_E) + D(P_E) \sim \tau_w$ is a necessary condition of eddy saturation. Furthermore, if the eddy energy dissipation is assumed to be represented by a fixed linear dissipation rate, i.e., $D(K_E) + D(P_E) = \lambda E$, $(1)$ is recovered. Physically, these conditions imply that eddy saturation is maintained by compensating the excess kinetic energy input associated with strengthened westerly through increase in the eddy energy sink.

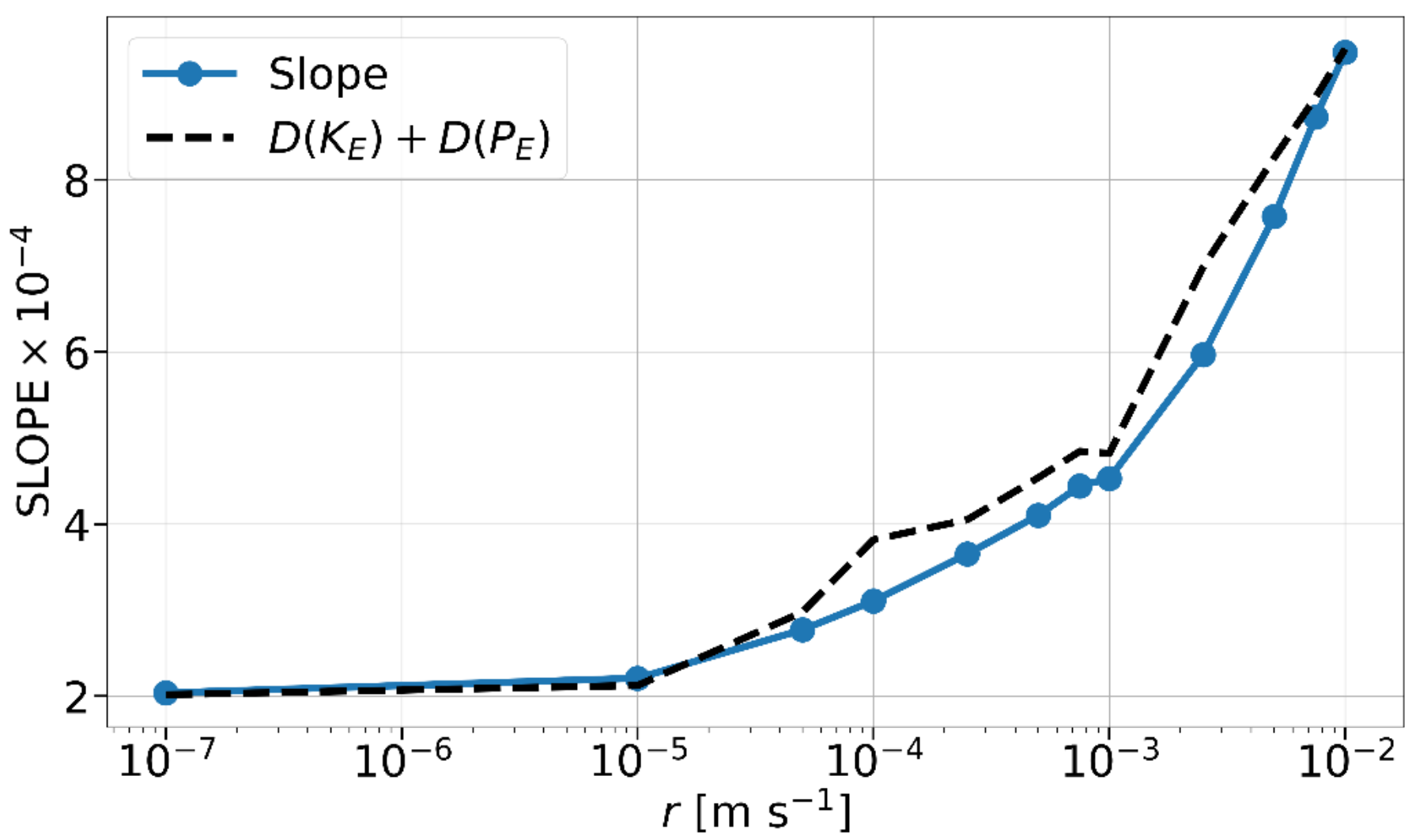

Figure 8 Curve of baroclinicity as a function of drag coefficients. Black dashed curve indicates a function proportional to eddy energy dissipation with a constant offset.

## 5. Summary and conclusion

In this study, we conducted a series of sensitivity analyses using an eddy-rich idealized reentrant channel model to revisit the frictional control mechanism of the ACC proposed by previous studies. By modifying the drag coefficient, we investigated the sensitivity of mean flow structures, eddy energy, the LEC, and baroclinicity to friction. Figure 2 shows that the flow becomes more sensitive to bottom topography as the bottom drag coefficient decreases because the mean flow becomes more barotropic (Figure 3). As a result, wind-driven gyre circulations are evident under low-drag conditions, while the zonal circumpolar component is dominant under medium- and high-drag conditions. Although standing meanders are evident in all regimes, their meridional deflections decrease with increasing drag coefficient. As shown in Figure 4, EKE is insensitive to bottom drag in the high-drag regime as suggested in previous studies. However, EKE except for the high-drag regime, EAPE, and total eddy energy are strongly dependent on the bottom drag, suggesting that a key ingredient of frictional control (1) with a constant tuning parameter is not valid in our configuration. The disagreement between (1) and numerical results suggests that the tuning parameter in the eddy geometric parameterization may need to vary with friction. By analyzing the LEC (Figure 6), it is found that energy pathways depend on the bottom drag coefficient. Under high-drag conditions, a large part of the wind energy is transferred to the APE reservoir, and the APE gain is primarily converted to EKE (Figure 7). By contrast, under medium- and low-drag conditions, the barotropic energy pathway became non-negligible owing to enhanced standing meanders (Figure 5). In these regimes, the mixed-instability nature is likely to weaken the correspondence between eddy energy and eddy interfacial form stress, which may be one of the reasons why (1) does not hold.

Our results suggest that the choice of drag coefficient may be important for realistic ocean general circulation model studies, since the shift in the BCR fraction occurs between $10^{-3}$ m s$^{-1}$ and $10^{-4}$ m s$^{-1}$ (Figure 7). In fact, different studies have reported varying contributions of BCR in the ACC. For example, some studies have shown that BCR is dominant in the ACC (Chen et al. 2014; Matsuta and Masumoto 2023; Matsuta et al. 2024), whereas others considered BTR to be non-negligible (Wu et al. 2017). Such inter-model differences in the LECs likely depend on the bottom drag parametrizations.

In section 4, Figure 8 demonstrates that the baroclinicity increases with drag coefficient and (2) is valid under all regimes, even though (1) is not. To explain this behavior, we generalized the frictional control framework without relying on any specific eddy parameterization. Instead of assuming that eddy energy can be scaled solely by the wind, i.e., condition (1), we focused on the energy diagram and directly characterized the BCR in terms of the wind forcing and the baroclinicity. Using this assumption, we obtained the generalized frictional control relation. The generalized frictional control (21) suggests that eddy energy dissipation controls baroclinicity, which supports implications of previous studies (Marshall et al. 2017; Mak et al. 2017, 2018). We showed that (21) is in good agreement with the results of our numerical experiments (Figure 8).

We conclude that the frictional control reasonably holds in stratified reentrant channels as long as the baroclinic energy pathway is dominant. Since the baroclinic energy pathway is also dominant in the realistic ACC (e.g., Chen et al. 2014; Matsuta and Masumoto 2023; Matsuta et al. 2024), the eddy energy dissipation is likely to regulate the ACC transport as suggested in previous studies (Mak et al. 2018, 2022a,b). To further understand the frictional control mechanism of the real ACC, future work must improve parametrizations of eddy energy dissipation. Although we assumed linear drag for simplicity, the dissipative processes in the real ACC are highly complex, being influenced by tides, topography, and stratification (e.g., Waterman et al. 2013; Naveira Garabato et al. 2013, 2016; Trossman et al. 2016; Yang et al. 2018; Klymak 2018; Hibiya 2021; Takahashi and Hibiya 2021; Klymak et al. 2021; Yang et al. 2023; Zhu et al. 2025). For example, Klymak (2018) and Klymak et al. (2021) suggested that baroclinicity in the ACC increases with increasing mesoscale topographic height, suggesting that topographic roughness alters eddy dissipation. Another study suggested that small-scale topography influences energy transfers and controls the strength of the large-scale jets (Zhu et al. 2025). Furthermore, Yang et al. (2023) suggested that linear drag does not represent the lee wave drag and they suspected that eddy saturation cannot be achieved if the lee wave generation is particularly vigorous. The interplays among eddy, standing meander and dissipative processes should be investigated in future work.

In this study, we have focused on baroclinicity as a mechanism that determines circumpolar transport. However, other mechanisms for frictional control have been proposed in previous studies. Some studies pointed out that the interaction between barotropic instability and standing meanders controls form drag (Constantinou and Young 2017; Constantinou 2018; Constantinou and Hogg 2019). Matsuta and Mitsudera (2024) also showed that increased nonlinearity of the gyre modifies the form drag, leading to a reduction in transport as friction decreases. Furthermore, we showed that Rossby wave radiation associated with barotropic instability possibly rectifies the barotropic transport (Matsuta et al., submitted to *JPO*), indicating that a parametrization of barotropic eddy flux may be necessary to simulate the barotropic dynamics of ACC (Eaves et al. 2025). Since these previous studies were conducted using unstratified models, future work should examine whether these barotropic mechanisms also operate in stratified models.

*Acknowledgments.*

Takuro Matsuta was supported by JSPS Grant-in-Aid for JSPS Fellows No. 22J00651 and JSPS Grant-in-Aid No. 24K17120. This work was supported in part by the Collaborative Research Program of Research Institute for Applied Mechanics, Kyushu University, and the Cooperative Research Activities of Collaborative Use of Computing Facility Program (JURCAOSCFG25-06) of Atmosphere and Ocean Research Institute, The University of Tokyo. This research was conducted using the FUJITSU Supercomputer PRIMEHPC FX1000 and FUJITSU Server PRIMERGY GX2570 (Wisteria/BDEC-01) at the Information Technology Center, The University of Tokyo.

*Data Availability Statement.*

MITgcm is available at https://mitgcm.readthedocs.io/en/latest/index.html.

## Appendix: Suppression of energy conversion from MKE to MAPE

To investigate the mechanism that suppressed the energy conversion from MKE to MAPE under low-drag conditions, we decompose the MAPE gain into the Ekman transport component and ageostrophic component in the ocean interior (Roquet et al. 2011):

$$C(K_M, P_M) = C^{(E)}(K_M, P_M) + C^{(I)}(K_M, P_M), \tag{A1}$$

where

$$C^{(E)}(K_M, P_M) = \iint \rho_0 g \, \mathbf{U}_{\mathrm{EK}} \cdot \nabla_h \overline{\eta} \, dxdy, \tag{A2}$$

and

$$C^{(I)}(K_M, P_M) = C(K_M, P_M) - C^{(E)}(K_M, P_M). \tag{A3}$$

Here, $\mathbf{U}_{\mathrm{EK}}$ is the Ekman transport and $\eta$ is the sea surface elevation. The term of (A2) can be interpreted as the work done by the pressure gradient associated with the sea surface height acting on the Ekman flow. Because the Ekman transport can be expressed as

$$\mathbf{U}_{\mathrm{EK}} = -\frac{1}{\rho_0 f} \hat{\mathbf{k}} \times \boldsymbol{\tau}_w, \tag{A4}$$

where $\hat{\mathbf{k}}$ is the upward unit vector and $\boldsymbol{\tau}_w$ is the steady wind stress vector, the integrand of (A2) is balanced by the wind stress work acting on the surface geostrophic flow:

$$\rho_0 g \, \mathbf{U}_{\mathrm{EK}} \cdot \nabla_h \overline{\eta} = \boldsymbol{\tau}_w \cdot \overline{\mathbf{u}}_g^{(0)}, \tag{A5}$$

where $\mathbf{u}_g^{(0)}$ is the surface geostrophic current.

Figure A1(a)-(c) indicate that the spacing of the sea surface height contours becomes wider as the drag coefficient decreases. Since $C^{(E)}(K_M, P_M)$ is determined by the meridional sea surface height gradient in our case, the MAPE gain decreases with decreasing drag coefficient. Indeed, areas of prominent positive $C^{(E)}(K_M, P_M)$ in HIGH are wider than those of MEDIUM and LOW cases. Although the values near the standing meander are enhanced in the MEDIUM and HIGH cases compared with the LOW case, as shown later, this enhancement is not large enough to change the ranking of the domain-integrated values.

The horizontal distribution of $C^{(I)}(K_M, P_M)$ (Figure A1(d)-(f)) also highlights a key difference caused by the drag coefficient. In LOW and MEDIUM, negative values of $C^{(I)}(K_M, P_M)$ are evident along the southward western boundary current, while the magnitudes of $C^{(I)}(K_M, P_M)$ are much smaller in HIGH. When friction is weak, the flow becomes more barotropic (Figure 3) and the influence of bottom topography becomes pronounced (Figure 2).

In MEDIUM and LOW, this results in the formation of a strong western boundary current. The pressure difference associated with the western boundary current accelerates its ageostrophic component, which corresponds to the conversion of the MAPE into the MKE. It should be noted that the magnitudes of both the positive and negative values increase as drag coefficient decreases. However, the integrated value is always negative, and its magnitude increases with decreasing drag coefficient as seen below. The emergence of patchy positive and negative structures may be related to the phase of standing meanders.

Figure A2 summarizes the dependency of $C^{(E)}(K_M, P_M)$ and $C^{(I)}(K_M, P_M)$ on the drag. The Ekman work gradually increases with the drag coefficient. The value is smaller than 60 GW in the low-drag regime, and increases to more than 80 GW at $r = 10^{-2}$ m s$^{-1}$. Figure A2(b) shows that in the low-drag regime, $C^{(I)}(K_M, P_M)$ is approximately $-15$ GW, but its magnitude rapidly decreases above $r = 10^{-4}$ m s$^{-1}$, reflecting a reduction in the barotropic mode fraction (Figure 3) and associated weakening of the western boundary current. In summary, a decrease in the drag coefficient reduces the efficiency of APE gain by Ekman flow due to reduced surface geostrophic current. Additionally, the pressure gradient associated with the western boundary current promotes conversion from MAPE to MKE, further reducing $C(K_M, P_M)$.

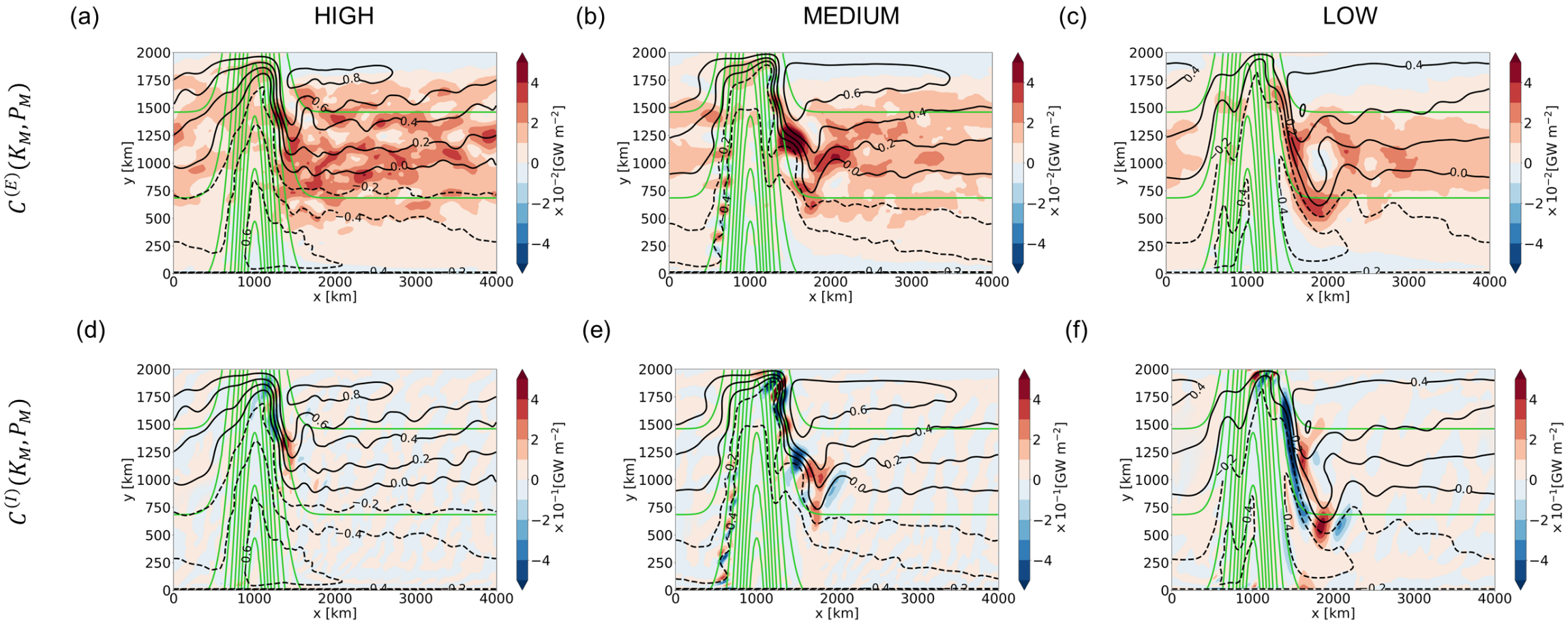

Figure A1. Horizontal distributions of $C^{(E)}(K_M, P_M)$ for (a) MEDIUM, (b) HIGH, and (c) LOW. Green contours are the same as those in Figure 2. Black contours are sea surface elevation contours with a contour interval of 0.2 m. Lower panel is same as the upper panel but for $C^{(I)}(K_M, P_M)$. Note that the color scale differs from that of the upper panel.

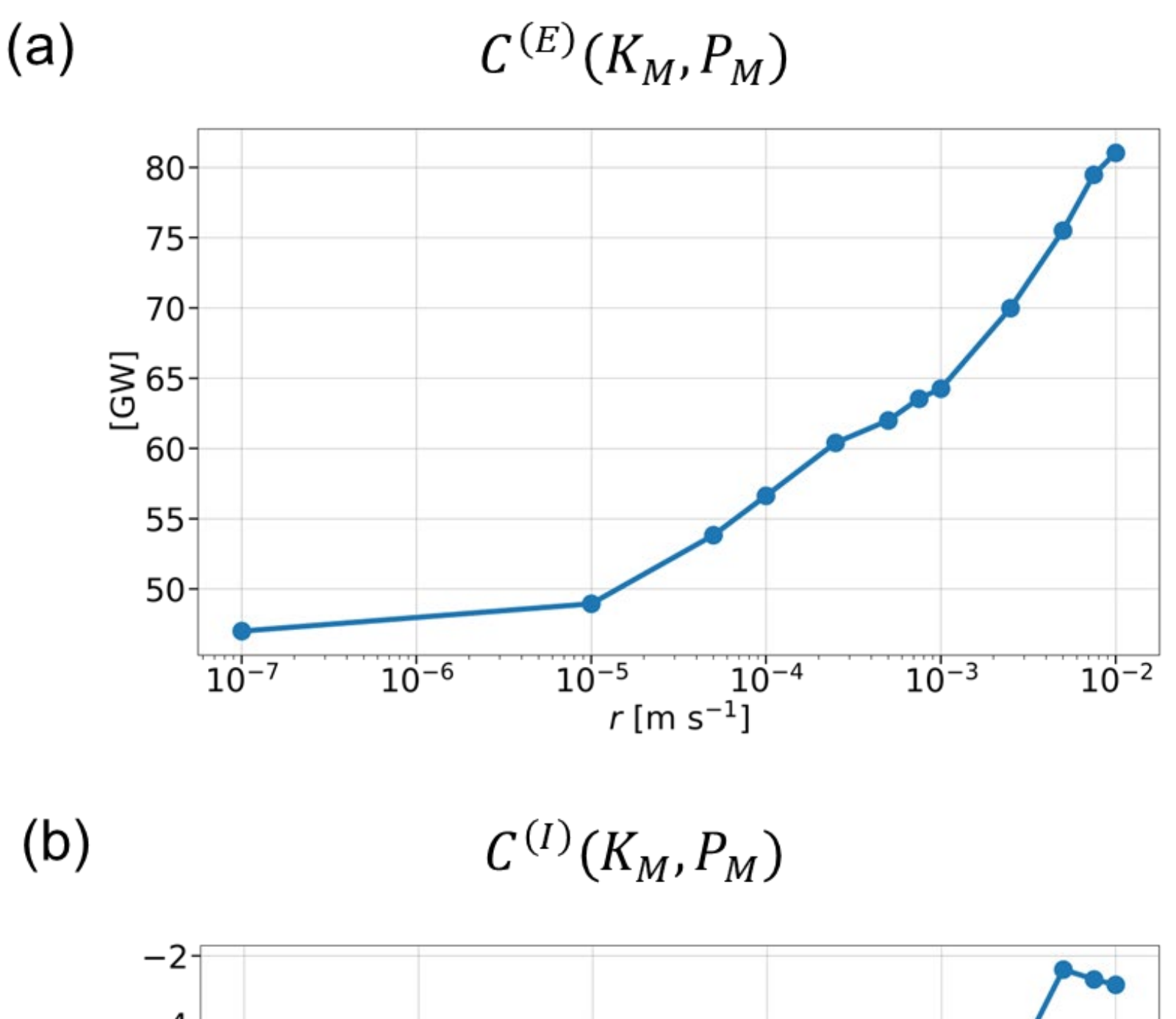

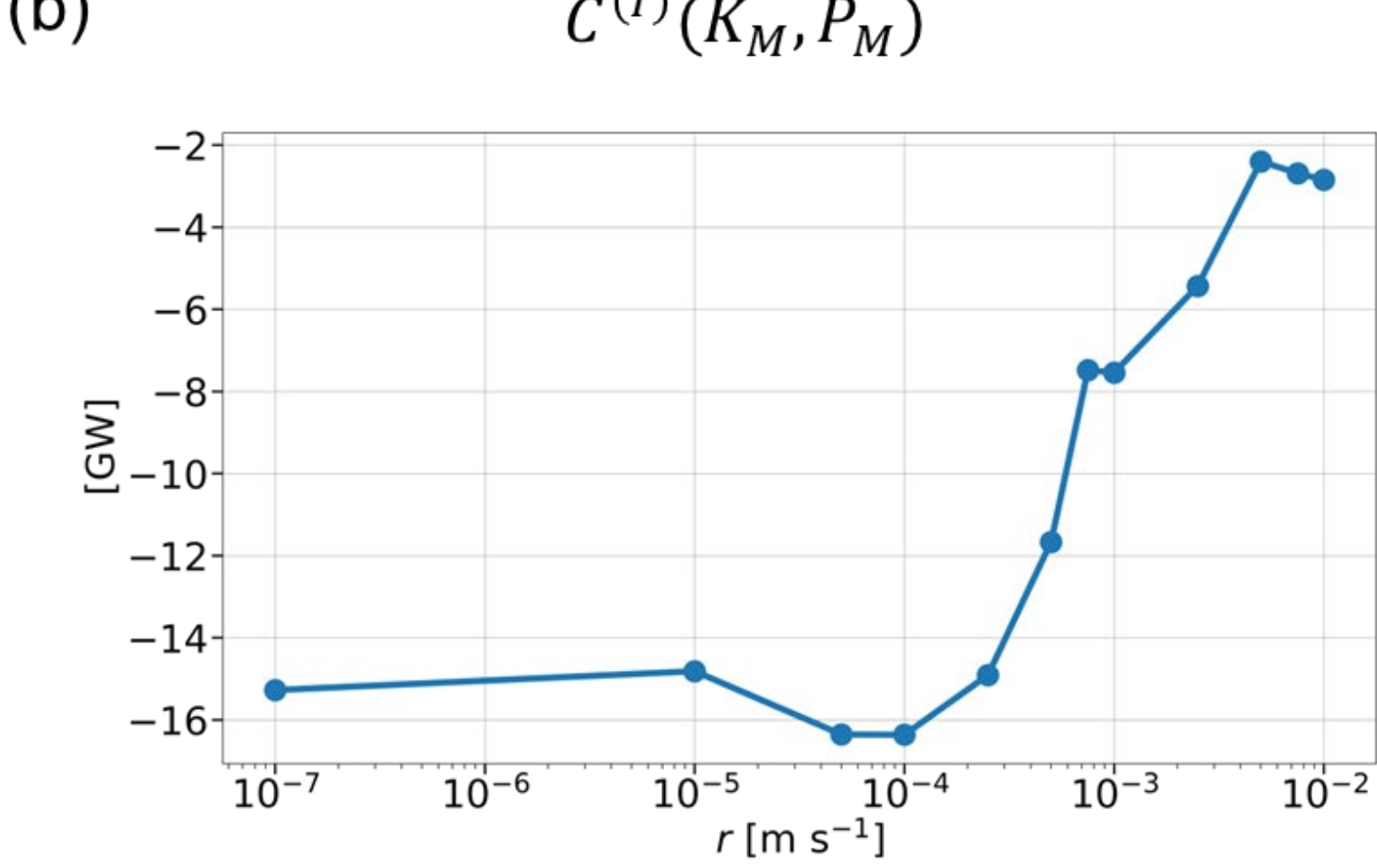

Figure A2 Curves of (a) $C^{(E)}(K_M, P_M)$ and (b) $C^{(I)}(K_M, P_M)$ as functions of drag coefficients.